\documentclass[12pt,aps]{revtex4}
\renewcommand{\baselinestretch}{1.5}
\begin{document}

\oddsidemargin = -0.52cm     
\textwidth = 16.96cm         
\topmargin = 0cm            
\textheight = 23.7cm        
\voffset = -0.54cm          
\paperwidth = 21cm          
\paperheight = 29.7cm       
\renewcommand{\baselinestretch}{1.3}    
\parskip=6pt


\title{Incoherent "Slow and Fast Light"}
\author{ \hskip50mm Zapasskii V.S. and Kozlov G.G. \newline Institute of Physics, St.-Petersburg State
University,
 St.-Petersburg, 198504 Russia. \newline {\it e}-mail:  gkozlov@photonics.phys.spbu.ru}

\begin{abstract}
We show experimentally that the effects of "slow and fast light" that
are considered to be caused by spectral hole-burning under conditions
 of coherent population oscillations (CPO) can be universally observed
 with incoherent light fields on objects with the pure-intensity nonlinearity,
 when such an interpretation is inapplicable.  As a light source, we used an incandescent
 lamp and as objects for study, a photochromic glass and a thermochromic coating.
 The response of the objects to intensity modulation of the incident light reproduced
 in all details the commonly accepted experimental evidences of the "light with a
 negative group velocity" and "ultraslow light". We come to conclusion that so far
 there are no experimental works providing evidence for real observation of the
 "CPO-based slow or fast light".
\end{abstract}
\maketitle
\section{INTRODUCTION}

Among the immense multitude of the effects of nonlinear optics, there may be singled out a
group of phenomena comprehensively described in terms of light intensity without using
explicitly the notion of the light field and its characteristics (frequency, spectral
width, etc.). In these cases, the relation between the output light intensity ($I_{out}$)
and intensity of the incident light ($I_{in}$) may be presented in the general form as
\begin{equation}
I_{out} = K(I_{in},t) I_{in},
\end{equation}

        A characteristic feature of these effects, which are not associated
        directly with nonlinear polarizability of the medium at optical frequencies,
        is their certain sluggishness controlled by the recovery time $\tau$ of the
        parameter $K$.   In the linear approximation, the dynamics of this recovery
        upon variation of the light intensity can be described by the simple differential equation
\begin{equation}
{dK\over dt} = {K_{eq}-K \over \tau},
\end{equation}
where $K_{eq}$ is the equilibrium value of the coefficient $K$ for the current intensity $I_{in}$.
 For relatively small variations of the light intensity $I_{in}$, the dependence $K_{eq}(I_{in})$
can be approximated by the linear expansion
\begin{equation}
K_{eq}(I_{in}) = K_0 + K_1 I_{in}, \hskip15mm K_1 I_{in}<< K_0,
\end{equation}
where the positive and negative signs of the coefficient $K_1$ correspond,
respectively, to the cases of super- and sublinear dependences \cite{1}.

Equations (1) - (3), in spite of their simplified form, allow one to describe all
the main features of response of such a nonlinear medium to variations of the incident
light intensity (Fig. 1) \cite{1}. In particular, temporal response of $I_{out}$ to stepwise
change of $I_{in}$ exhibits exponential relaxation to its equilibrium value (Fig. 1a);
the frequency dependence of the response ($I_{out}$ ) to a weak intensity modulation of
$I_{int}$ shows a Lorentzian feature with a half-width $\sim 1/\tau$ peaked at zero frequency (Fig. 1b);
and the appropriate frequency dependence of the modulation phase of $I_{out}$  exhibits maximum
delay (positive or negative depending on the sign of $K_1$)  at the frequency $\sim 1/\tau$  (Fig. 1c).
A pulse of the light modulation, in the general case, appears to be distorted. However,
at low modulation frequencies, the absolute value of the time delay becomes frequency-independent,
and, as a result, a sufficiently long smooth pulse experiences a pure shift with no
reshaping (Fig. 1d). The sign of this shift may be also either positive or negative
depending on the sign of  $K_1$ (for more detail see \cite{1}).
One can notice that solutions of Eqs. (1) - (3) (Fig. 1) demonstrate alterations in
the intensity spectrum of the light resulting from interaction with the nonlinear medium.
For a spectroscopist, the easiest way to make it sure is to make use of the light with
a "white" intensity spectrum, i.e., the light with its intensity modulated by "white" noise).
Then, at the exit of the medium, the intensity spectrum of the light, initially flat, will
display a peak (or dip) with a maximum at zero frequency and with the width $\sim 1/\tau$. Usually,
however, the frequency dependences of this kind are obtained using the light with a
"monochromatic" intensity spectrum: the intensity $I_{in}$ is subjected to a weak
harmonic modulation, with the amplitude and phase of $I_{out}$  being measured at the
exit of the sample at different frequencies.

The optical media with the above type of nonlinearity are usually referred to as
saturable absorbers. The superlinear and sublinear dependences \cite{1} correspond to
two types of  saturable absorbers, namely to the bleachable and so-called inverse
absorbers. In the simplest cases, the optical nonlinearity \cite{1} is the result of
light-induced changes in populations of eigenstates of the system, with the characteristic
recovery time $\tau$ being determined by the population relaxation rate. Sometimes,
the nonlinearity of this kind arises in optical systems capable of transforming
nonlinear variations of geometrical or polarization characteristics of the beam
into variations of the intensity signal, e.g., upon formation of the Kerr-type lens
in the systems with vignetting elements or upon light-induced polarization plane rotation
in the systems with polarizing elements. Such systems are sometimes referred to as artificial
saturable absorbers. The intensity-type nonlinearities can be also observed in reflection,
in scattering, in atomic systems under conditions of optical orientation, in impurity
crystals and glasses, in bulk and low-dimensional semiconductors, upon photochemical
and photoconformation processes, upon light-induces heating of the medium, and so on,
and so forth. Moreover, Eqs. (1) - (3) containing no carrier frequency and, generally,
not implying its presence have no optical specifics and, evidently, may describe a
much wider class of nonlinear physical problems.  In \cite{2}, in particular, this circumstance
has been demonstrated by an example of an electric circuit with a nonlinear resistor.

All the aforesaid actually refers to well-known effects of nonlinear optics (or, better to
say, of nonlinear physics). In particular, the effect of the light pulse delay in a
saturable absorber is known for more that 40 years \cite{3}, and the media with intensity
nonlinearity are widely used starting from 60s in passive systems of Q-switching and mode-locking \cite{4,5}.
In the end of the 20th century, however, the above effects of retarded response of a
saturable absorber have been rediscovered \cite{6,7} and, without mentioning the known
mechanism of their origination, were ascribed to the light-induced changes in the
group velocity of light in saturable absorbers.
As is known, the group velocity of light pulse with the carrier frequency $\omega_0$
  in a dispersive medium
\begin{equation}
V_{gr} = {c\over n + \omega_0 dn/d\omega}
\end{equation}
 depends on steepness of the refractive index spectral dependence $dn/d\omega$ and, in principle,
may strongly differ from the "phase" velocity of light $c/n$. In the proposed interpretation,
the great value of the dispersion term in Eq. (4) was considered to be a consequence of a
narrow dip in the spectrum of the saturable absorber burnt by the  monochromatic pump wave
due to the effect of coherent population oscillations (CPO) \cite{8}. In other words, the narrow
peak observed at low frequencies in the intensity spectrum of the light transmitted by the
nonlinear medium was now regarded as lying in the optical spectrum of the medium.  Evidently,
the new interpretation of the pulse delay in a saturable absorber invoking the effects of
propagation substantially differed from the previous interpretation and, in the general case,
could not be equivalent to the latter.
        Thereafter, this new idea, in spite of its being seriously criticized (see \cite{1,9,10}),
        has gained a wide popularity. There have been published dozens of papers on observation
        of such a "slow" and "fast" light, and the CPO effect turned into one of the basic
        technologies of this field of optics (see, e.g., \cite{11,12} and references therein).
        Such an interpretation of the effects of intensity nonlinearity proved to be viable
        probably because all the known experiments in this field were performed using laser
        sources, thus providing a preferential optical frequency needed to position the narrow
        spectral dip in its vicinity. In addition, all the theoretical papers were also based on
        the model of monochromatic electromagnetic fields.
In this paper, we show experimentally that the effects of the so-called "slow" and "fast" light
are universally observed in the incoherent light on non-laser sources on arbitrary objects with
the intensity-type nonlinearity. As the light source, we used an incandescent lamp, with the object
for study being a photochromic glass and a decorative thermochromic coating. Both objects were
characterized by rather long relaxation times ($\sim$ 10 s), which also provided a relatively low
threshold of the optical nonlinearity. The time delays of intensity modulation of the light
after interaction with the medium demonstrated what is usually considered as the "light with
negative group velocity" and "ultraslow light" though the CPO-based model of the group velocity
modification, in this case, cannot be valid.

\section{"Negative group velocity of light" in photochromic sun glasses}

The first object for study was a photochromic spectacle glass (Opticoelektron group JSC - Panagyurishte)
that our optician has managed to find in the drawer of his desk. In the preliminary experiments,
we studied dynamics of photochromic darkening of the glass at different power densities of the irradiation.
The experiments were performed in the pump-probe configuration using two light sources - a halogen
incandescent lamp ($P$ = 90W for the current 7.5A) as the pump and a laser diode ($P \sim 1$ mW, $\lambda\sim$650 nm)
 as a probe (Fig. 2).
The pump light was focused onto the glass as a spot $\sim$3x3 mm$^2$ in size. The probe laser beam passed
through the sample practically in the opposite direction and initially was centered on the spot of
the pump and then, when measuring, was translationally modulated at an audio frequency with the aid
of a prism fastened to membrane of a loudspeaker (Fig. 2). Cross section of the laser beam ($\sim$ 3 mm$^2$)
was substantially smaller than the size of the spot, while the amplitude of its spatial displacement
upon modulation was slightly larger than the spot. As a result, due to the photoinduced darkening of
the glass, the intensity of the probe beam transmitted through the sample revealed modulation,
proportional to the darkening, synchronized with vibrations of the membrane. The above modulation
signal was lock-in amplified and recorded as a function of time. The photodetector was placed in
the focal plane of the lens which allowed us to eliminate the parasitic modulation caused by spatial
modulation of the spot over the photodetector.
        The current of the incandescent lamp was controlled by a special generator with the lower
        frequency bound of 0.001 Hz.
Upon stepwise change of the pump light intensity, the time-dependent signal of the probe beam showed
dynamics of recovery of the glass transmission. These measurements were performed under conditions
 of square-wave modulation of the lamp current at frequencies of about a few hundredths of Hz with
the degree of intensity modulation of about 20 - 30 $\%$.
We have found that the relaxation dynamics strongly depended on the irradiation intensity. At low
levels of illumination, the relaxation times were rather long (hundreds of seconds), then, as the
pump intensity increased, they shortened to tens of seconds and then the situation became unstable:
both the relaxation time and even the sign of the photochromism showed a long-term drift in the range
of tens of minutes. The latter region was highly inconvenient for our measurements, and we have chosen
the range of power densities where the relaxation times were relatively short. Figure 3
shows experimental time dependence of transmission of the glass sample
 upon square-wave modulation
 of the lamp current  in the vicinity of 4.5 A. The relaxation time of the induced absorption,
under these conditions, was around 10 s, though, strictly speaking, the relaxation process was
not single-exponential.

The experiments were performed in a standard way, as in most experiments on "slow" and "fast"
light. Intensity of the pump light incident on the sample was subjected to a weak harmonic modulation,
whose amplitude and phase were measured at the exit of the sample at different modulation frequencies.
The results of the measurements are shown in Fig. 4. As expected for the inverse saturable absorber,
the amplitude spectrum of the signal exhibited a dip in the range of low frequencies (rather than a
peak typical for a bleachable absorber), while the phase spectrum corresponded to {\it advancement} of the
output signal with respect to that at the entrance. In view of the non-exponential character of the
relaxation mentioned above, the shape of the experimental frequency dependences did not exactly
correspond to predictions of the simplest model \cite{1}, which, for our purposes, was of no importance.
Figure 5 shows schematically mutual position of normalized curves of the light modulation at the
entrance and at the exit of the sample for the modulation frequency 0.001 Hz. Temporal advancement
of the output light was, in this case, $\sim$ 3 s. If we assume, as it is accepted in the experiments on
"slow" and "fast" light, that the observed phase delay is related to a change of the light group
velocity in the medium, then we will have to acknowledge that the group velocity, in this case,
is negative and equals - 0.67 mm/s. Now, however, the possibility of such an interpretation is
excluded, because there is no preferential frequency in the optical spectrum of the light source
near which the appropriate anti-hole (with a width of about 0.1 Hz) is supposed to be burnt, and
therefore the delay cannot be ascribed to a change in the group velocity of light in the sense of
 Eq. (4). It is evident that the terms"superluminal light" or "the light with negative group
 velocity" are also inappropriate here.

\section{"Ultraslow light" in a thermochromic coating}

As the second object for study, we used a fragment of a cup with a thermochromic decorative coating
that changed its color upon heating (in this particular case, the initially red coating was noticeably
bleached upon heating to 40$^o$ - 50$^o$).
Dynamics of response of the sample to variations of the light intensity was studied using the same
incandescent lamp with the same experimental arrangement, with the only difference that now we detected
the light scattered by the surface of the sample (rather than transmitted light). The photodetector
was placed approximately at the angle of specular reflection, where the light intensity scattered by
the glossy surface of the coating was the greatest. In front of the detector, we placed an optical
filter transmitting blue-green part of the visible spectrum, which provided a higher
  contrast of the detected intensity nonlinearity. With
increasing intensity of the incident light, the temperature of the sample was growing, the coating
was getting bleached, and the effective reflectivity increased. Figure 6 shows the experimental
dependence of equilibrium intensity of the reflected light on that of the incident light. The
dependence was recorded under conditions of ultimately slow scanning of the lamp current
(during ~ 10 min). This is the basic dependence describing the intensity-type nonlinearity of the
nonlinear medium [see Eq. (1)], which is, in this case, a saturable reflector. The growing reflectivity
with increasing light intensity corresponded to the positive sign of the coefficient $K_1$ in Eq. (1).
For such a high level of nonlinearity, dynamics of the relaxation behavior of the sample could be easily
observed directly in the reflected light upon
 square-wave modulation of the incident light intensity
 (Fig. 7). The recovery time of the sample, in our experimental conditions, lied in the range of 15 s.
The experimental frequency dependence of the amplitude and phase of the reflected light intensity
modulation are shown in Fig. 8. For normalization of the amplitude and phase of the reflected signal,
we used an additional photodetector illuminated by the light split from the incident beam.
The frequency behavior of the response, in this case, qualitatively coincided with what is usually
observed for a bleachable absorber under conditions of laser excitation. The modulation amplitude
shows a growth in the vicinity of zero frequencies (Fig. 8a), while the phase shift (Fig. 8b)
reveals a peak in the vicinity of the frequency  approximately corresponding to the inverse
recovery time of the nonlinear medium, and, by its sign, to positive delay of the modulation
with respect to that of the incident light. The value of this delay, in the limit of low
frequencies, was about 2.5 s. In this case, however, this delay not only cannot be ascribed
to the group velocity reduction, but, in addition, the group velocity cannot be estimated in a
sensible way, though the experiment evidently may be performed, with the same results, using
laser source

\section{Conclusions}

In this paper, we have shown experimentally that the effects of the light pulse delay (or phase
delay of the light intensity modulation) are inevitable in any medium with the intensity-type
nonlinearity and, by themselves, do not evidence in favor of any changes in the light group velocity.

        The essence of all the effects of this kind is reduced to
        changes in {\it intensity spectrum}
        of the light interacting with the nonlinear medium and, correspondingly, to changes in
        temporal behavior of its intensity. In the particular case of a sufficiently long pulse
        (when the main part of its intensity spectrum falls into the region of linear dependence
        of the intensity modulation phase on modulation frequency), the pulse distortion is reduced
        to its pure shift in time. In this case, the spectral composition of the optical field is
        important only to the extent of its influence  on
         parameters of Eqs. (1) - (3).

        For this reason, all the experimental results on the delay or advancement of the light
        pulse propagating through a saturable absorber (starting from the paper \cite{6}) are natural
        for the media with intensity nonlinearity and do not require, for their interpretation,
        invoking the effects of the light group velocity changes
        In principle, the described variations in the intensity spectrum of the light interacting
        with a saturable medium should also be revealed in fine changes of the phases and amplitudes
        of the optical spectral components (this is, in fact, the essence of the CPO effect). However,
        the contribution of such processes into changes of the light group velocity can be revealed
        only in purposely designed experiments, which, to the best of our knowledge, so far has not
        been performed. Even in the experimental work \cite{13}, where the steep
        refractive index dispersion under conditions of the CPO
        effect was convincingly demonstrated, no direct measurements
        of the probe pulse delay in the presence of the pump have been
        made and compared with calculations. Such measurements make
        sense for two reasons. First, the CPO-related dip in the
        absorption spectrum is not the same for monochromatic probe
        and for the probe with a continuous spectrum (because of interaction
        between the spectral components of the probe symmetric with respect
        to the pump). In particular, as was pointed in \cite{1}, for the probe
        light with certain phase correlations between its "mirror" components,
        the dip can be unobservable. And, second, it is known that the pulse
        delay can be also observed for the probe pulse wavelength shifted far
        away from that of the pump  \cite{14}. This is why it would be important to
        evaluate the "resonant" CPO-based contribution to the delay of optical
        pulse in a saturable absorber.

        In any case, we are not aware of any experimental observations on the "CPO-based slow light"
        that cannot be interpreted in the framework of the simplest model of intensity nonlinearity.
        The same conclusion was made in \cite{10} on the basis of a comprehensive analysis of many published
        experiments.
        In spite of the fact that the "CPO-based slow light"is nowadays one of the basic technologies
        of "slow light", in our opinion, there is no evidence of its real experimental observation
        and no grounds to consider its existence proven.

\newpage
\hskip40mm {\bf FIGURE CAPTIONS}

FIG.1 Regularities of  optical response of the medium with the
intensity-type nonlinearity. a - response to a stepwise change of
the incident light intensity; b and c - frequency dependences of
the amplitude and phase of a sine-modulated beam at the exit of
the medium; d - temporal shift of a smooth  pulse of intensity
modulation. Curves 1 and 2 correspond to the cases of superlinear
and sublinear dependence (1).

FIG.2 Experimental setup for measuring dynamics of darkening in
the photochromic glass. 1 - incandescent lamp, 2 - focusing lenses,
3 - photochromic glass, 4 - photodetectors, 5 - laser diod,
6 - vibrating prism, 7 - data processing system, 8 - power supply,
9 - controlling low-frequency generator.

FIG.3 Experimental time dependence of the photochromic glass darkening
 (a) induced by the square-wave modulated pump light (b). 

FIG.4 Frequency dependences of the amplitude (a) and
phase (b) of the sine-modulated light transmitted through the photochromic glass.

FIG.5 A sketch of time advancement of the sine- modulated
light transmitted through the photochromic glass.

FIG.6 Experimental dependence $I _{out}(I_{in})$ for the thermochromic coating.

FIG.7 Time dependence of intensity of the light reflected from the
thermochromic coating for the square-wave modulated incident light of the incandescent lamp.

FIG.8 Frequency dependences of the amplitude (a) and phase
(b) of the sine-modulated light reflected from the thermochromic coating.

\end{document}